\documentclass[journal,10pt]{IEEEtran}
\IEEEoverridecommandlockouts
\usepackage{cite}
\usepackage{soul}
\usepackage[numbers,sort&compress]{natbib}
\usepackage{amsmath,amssymb,amsfonts}
\usepackage{algorithm,algpseudocode}
\usepackage{graphicx}
\floatname{algorithm}{Algorithm}
\usepackage{titlesec}
\makeatletter
\newcommand*{\rom}[1]{\expandafter\@slowromancap\romannumeral #1@}
\makeatother
\usepackage{hyperref}
\usepackage{caption}
\usepackage{fixltx2e}
\usepackage{tabularx}
\usepackage{makecell}
\hypersetup{
colorlinks=false,    % set true if you want colored links
linkbordercolor={1 1 1}, % set to white {1 1 1} or {0 0 0} for black
citebordercolor={1 1 1}, % set to white for citation links
urlbordercolor={1 1 1}   % set to white for URL links
}

\usepackage[table]{xcolor}
\usepackage{xcolor}
\usepackage{float}

\usepackage{tikz}

\usepackage{listings}
\definecolor{codegreen}{rgb}{0,0.6,0}
\definecolor{codegray}{rgb}{0.5,0.5,0.5}
\definecolor{codepurple}{rgb}{0.58,0,0.82}
\definecolor{backcolour}{rgb}{0.92,0.92,0.92}
\usepackage{textcomp}
\newcommand{\RNum}[1]{\uppercase\expandafter{\romannumeral #1\relax}}
\usepackage{xcolor}
\usepackage{listings}
\def\BibTeX{{\rm B\kern-.05em{\sc i\kern-.025em b}\kern-.08em
	T\kern-.1667em\lower.7ex\hbox{E}\kern-.125emX}}
	
\begin{document}

\title{LAAG-RV: LLM Assisted Assertion Generation\\ for RTL Design Verification}

\author{\IEEEauthorblockN{Karthik Maddala, Bhabesh Mali, Chandan Karfa}\\
	\IEEEauthorblockA{Indian Institute of Technology Guwahati, India\\
		\ \{k.maddala, m.bhabesh, ckarfa\}@iitg.ac.in}}

\maketitle

\begin{abstract}
	Writing SystemVerilog Assertions (SVA) is an important but complex step in verifying Register Transfer Level (RTL) designs. Conventionally, experts need to understand the design specifications and write the SVA assertions, which is time-consuming and error-prone. However, with the recent advancement of transformer models, the Large Language Models (LLMs) assisted assertion generation for design verification is gaining interest in recent times. Motivated by this, we proposed a novel LLM-based framework, LAAG-RV, to generate SVA from the natural language specifications of the design. Our framework provides a one-time Verilog loop for signal synchronization in the generated SVA to improve the generated assertion quality. For our experiments, we created a custom LLM based on OpenAI GPT-4. Furthermore, we developed test cases to validate the LLM-generated assertions. Initial observations show that some generated assertions contain issues and did not pass all the test cases. However, by iteratively prompting the LLMs using carefully crafted manual prompts derived from test case failures in a simulator, the framework can generate correct SVAs. Our results on OpenTitan designs demonstrate that LLMs significantly simplify the process of generating assertions, making it efficient and less error-prone.
\end{abstract}


\section{Introduction}
In Electronic Design Automation (EDA) flow, design verification is one of the important phases, ensuring that a design meets its specifications and functions correctly under all the expected conditions \cite{wang2009electronic}. Property-based verification is one of the key approaches in this phase, where specific properties are written to describe the expected behavior of the design. These properties serve as the formal statements that must hold true for the design to be considered correct. The primary purpose of writing properties is to specify and verify the functional and temporal aspects of the hardware design. These properties are converted into System Verilog assertions during implementation to facilitate their integration into the verification environment. Property verification has two broad methodologies, namely dynamic Assertion Based Verification (ABV) and Formal Property Verification (FPV) \cite{dasgupta2006roadmap}.

The FPV utilizes a model-checking tool. A model-checking algorithm has two main inputs: a formal property and a finite state machine representing the implementation. The role of the algorithm is to search all possible paths of the state machine for a path trace which refutes a property, which is reported as the counter-example. Otherwise, the model checker asserts that all the properties hold in the implementation.
In the dynamic ABV approach, on the other hand, we first write a test bench to provide inputs into our implementation.  
The design is simulated on the test bench. The assertion checker reads the signals and monitors the status of the properties and reports if any failures occur. The primary advantage of ABV is that it builds over a traditional simulation framework with nominal additional efforts and it does not have any capacity concern like FPV.
Given the increasing complexity of modern designs, our focus is on dynamic ABV. Dynamic ABV uses assertions, which are derived from its specification by the verification experts.  The properties used to describe the behavior of a hardware design must adhere to three fundamental principles known as the ``Three C's"- Correctness, Consistency, and Completeness.

``Correctness" of a property ensures that the property accurately captures the intended behavior of the design. It implies that the properties must accurately reflect the functional requirements and constraints of the design. ``Consistency" refers to the set of properties that don't contradict each other. In a verification scenario, consistency ensures that the property describes the non-contradictory behavior of the design across all possible scenarios. ``Completeness" refers to the properties that must cover all relevant conditions and corner cases in the design. The verification process must not miss any critical aspects of the system's behavior. For example, in a bus arbiter,

\begin{lstlisting}
	property p_correctness;
	@(posedge clk)
	disable iff (!rst)
	(grant == 1) |-> !($rose(grant) && bus_req);
	endproperty
\end{lstlisting}

represents a correct assertion that ensures if a bus request is granted to a master, no other master should be granted the bus at the same time.

\begin{lstlisting}
	// A granted bus request should be acknowledged within 2 clock cycles.
	property p_consistency1;
	@(posedge clk)
	disable iff (!rst)
	(grant == 1 && ack != 1) |-> ##[1:2] ack == 1;
	endproperty
	
	// No two masters should be granted the bus simultaneously.
	property p_consistency2;
	@(posedge clk)
	disable iff (!rst)
	(grant == 1) |-> !($rose(grant) && bus_req);
	endproperty
\end{lstlisting}

represents consistent properties where property 1 ensures a granted bus request should be acknowledged within 2 clock cycles. Property 2 ensures no two masters should be granted the bus simultaneously. On the other hand, ``Completeness" ensures the coverage of all the design aspects.

As the complexity of the design increases, so does its verification time. This rising complexity drives our motivation to use advanced technologies to enhance the verification process, especially Large Language Models (LLMs). LLMs has advanced to assist individuals in resolving complex challenges in various domains.
The hardware design industry is also using these models to accelerate the EDA flow \cite{chipchatMLCAD}. By integrating LLMs into the dynamic ABV process, we aim to streamline the creation of verification environments. This integration has the potential to significantly reduce the verification burden and improve the efficiency and effectiveness of the design verification process.

This work analyzes how assertions can be generated from natural language specifications, automatically, to assist in the ABV process. We investigate to answer two primary questions through our work: {\textit{ReQ1) How efficient are LLMs in producing SVA from natural language specifications?}} \textit{ReQ2)  Are the generated assertions functionally correct when integrated into the design verification environment?}

Our contributions are as follows:
\textcolor{black}{\begin{itemize}
		\item Development of a framework for investigating the application of LLMs in generating assertions for ABV.
		\item Introducing an effective manual prompting designed to provide relevant data to LLMs, enhancing their ability to generate precise assertions.
		\item Evaluation of LLM performance in producing correct assertions for a set of OpenTitan designs.
\end{itemize}}

The paper is organized as follows: Section \rom{2} discusses the background and related works. Section \rom{3} presents the system design and case study of the framework LAAG-RV. The experimental results and discussion are presented in Section \rom{4}. Finally, Section \rom{5} concludes the paper.
\section{related works and Background}

\subsection{LLM Overview}
The evolution from Probabilistic or Statistical Language Models (PLMs/SLMs) to Large Language Models (LLMs) represents a major advancement in the capabilities of natural language processing and understanding, as detailed in a variety of studies \cite{jelinek1998statistical, zhao2023survey}
These LLMs have evolved into valuable tools for human assistants, transforming fields such as medicine \cite{thirunavukarasu2023large, gilbert2023large}
and materials science \cite{jablonka202314, rubungo2023llm}. They assist researchers in managing large datasets, crafting innovative experimental designs, and formulating novel hypotheses.

LLMs are specific Transformer-based Language models where the training of the models uses large volume of data in text format, extracted from different sources such as web-books, websites and PDFs, to generate the weights parameters. LLMs can be categorized into two groups based on their application: \textbf{Domain-specific LLMs (ds-LLMs)} and \textbf{General Domain LLMs (gd-LLMs)} \cite{wang2023survey}. ds-LLMs are particularly used for a specific purpose in a specific area of interest. Examples include FoodGPT \cite{qi2023foodgpt} and FinGPT\cite{yang2023fingpt}. While gd-LLMs are trained on various resources across different domains and origins, enabling their application to a wide variety of queries. Some gd-LLMs are Llama 3 \cite{Llama3} and Gemini 1.5 \cite{reid2024gemini}. The training of LLMs on large datasets incurs significant costs. Therefore, refining pre-trained LLMs using fine-tuning is getting crucial to simplify the training process. Fine-tuning an LLM involves a supervised procedure, aimed at modifying the model parameters according to the specific task requirements \cite{zhang2023balancing}. Examples of fine-tuned LLMs include Star Coder \cite{li2023starcoder} and CodeLlama \cite{roziere2023code}.

\subsection{LLM-Driven Automatic Assertion Generation}

LLMs have advanced capabilities across various domains and tasks, such as generating source code \cite{chen2021evaluating}, automating software testing \cite{deng2023large, xia2023universal}, and correcting programs \cite{xia2022less, jiang2023impact}. They are particularly gaining popularity in automatic assertion generation from natural language specifications \cite{orenesvera2023using}. A notable study demonstrated that LLMs could generate assertions directly from RTL without any additional information. Although there were some initial errors, with appropriate guidance, LLMs were able to produce accurate and comprehensive assertions \cite{kande2023llmassisted}. Another research used comments from security assertions provided by engineers, deploying OpenAI's Codex to generate assertions. They produced over 75,600 assertions under different conditions and found that LLMs successfully generated correct assertions about 4.53\% of the time, with detailed comments and context-enhancing accuracy.

In the hardware security field, LLMs are increasingly used for security assertion generation as part of formal property verification. This helps detect vulnerabilities at the pre-silicon stage of development, allowing for early correction of security flaws in the design process \cite{witharana2023automated}. The work in \cite{mali2024chiraag} uses formatting of natural language specifications by providing certain labels in the JavaScript Object Notation (JSON), and RTL implementation is not provided to the LLMs to generate SVA. In contrast, we have provided a one-time verilog implementation loop to make sure we obtain refined SVA in a few iterations and generate correct SVA.

\section{System Design Methodology and Case Study}

This section provides an overview of our framework, LAAG-RV, shown in Fig. \ref{fig: flow_diagram}, which is used to generate SVA using LLM. LAAG-RV involves giving an initial description of the design to LLM and then prompting LLMs to generate SVA and further refine those generated assertions based on the issues. Each of the stages involved is described below:

\subsection{Specification for the LLM}
This phase involves the collection of detailed descriptions of the designs provided in the OpenTitan Repository \cite{Opentitan}. Each of the designs has multiple modules in the repository. Each design has multiple contexts, including a block diagram, description, register details, and Verilog Code. This description is given to the LLM as the initial prompt. However, we have only focused on the basic understandable details of the designs, excluding information about registers, Verilog implementation, etc, which depicts the client's perspective. 

\subsection{Prompting and Training LLMs}
A specific characteristic of LLMs is that it can generate different outputs for the same input. To maintain consistency, we performed all our experiments within a stable chat environment. We have used a custom LLM to automate the assertion generation process.

\subsubsection{Custom LLM}
We developed a custom GPT4 environment by enabling the functionality of GPT-4 along with Code-Interpreter abilities. This custom GPT4 was fed with domain-specific knowledge \cite{kande2023llmassisted, srikumar2023fast,orenesvera2023using, dasgupta2006roadmap,witharana2023automated}, providing the initial cognition ability and domain-specific knowledge.

\begin{figure}[!t]
	\centering 
	% \label{fig_11}
	\includegraphics[width=1.0\linewidth]{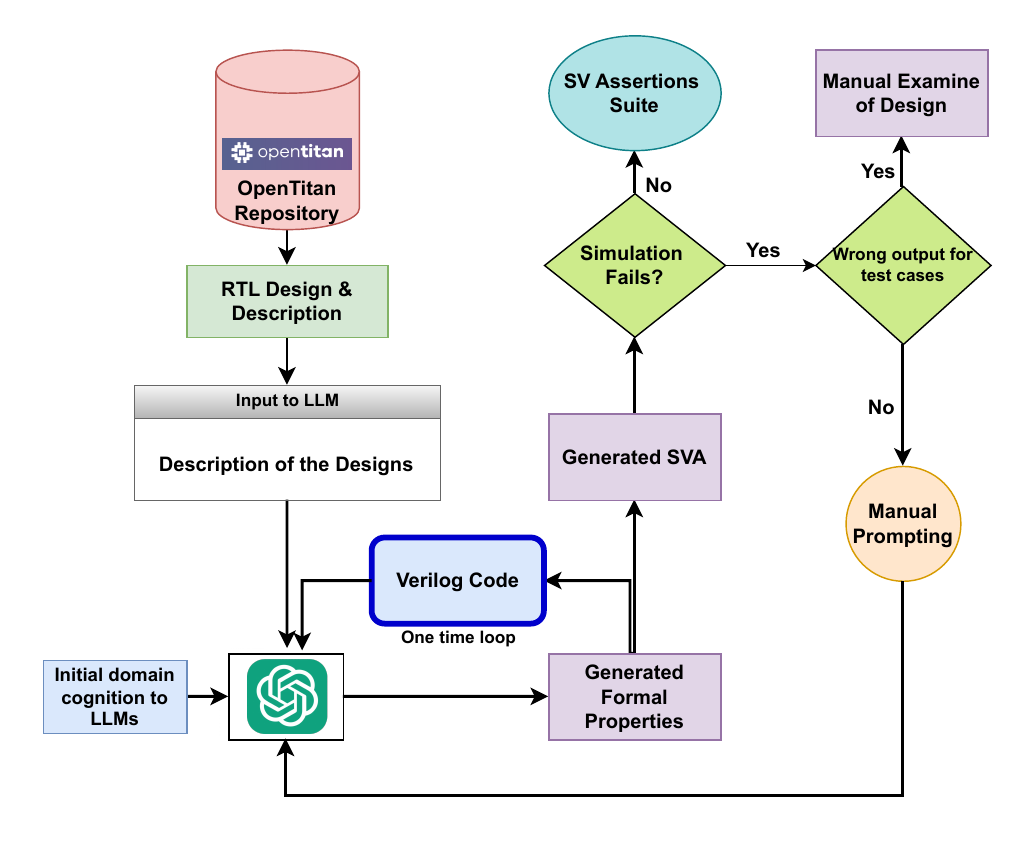}
	\caption{SVA Generation flow of LAAG-RV}
	\label{fig: flow_diagram}
\end{figure}

\subsubsection{Design-Specific Prompts to LLMs}

Initially, the design specification in natural language were inputted to the custom LLM as a starting prompt. The specification detailed the key functional requirements, generic architecture, and operational parameters of the designs. Additionally, we included the block diagram for each design in their respective chat environment, seeking a thorough understanding of the system's information flow.

\subsection{Signal Synchronization with Implementation}
Initially, the LLM is provided with schematic details, an overview of interfacing and is prompted to generate the properties. Subsequently, the LLM is asked to give the details of each generated property in English. This is done to understand if the generated properties are meaningful, i.e., semantically correct. Next in the  \textit{``One-time loop"}  in Fig. \ref{fig: flow_diagram}, an RTL code of the formal specification is supplied to the LLM from the OpenTitan Repository \cite{Opentitan}. \textcolor{black}{As the specification description doesn't contain any information about signal names, the RTL code is provided as a prompt to ensure signal names and other variables in the SVA} are synchronized with those specified in the RTL code.
This step assists in developing the test cases and performing verification. The LLM generates the raw SVAs and their English descriptions, which have yet to be verified.

\subsection{Error-Specific Prompts}
After generating initial raw assertions using LLM, we developed test cases to verify the generated SVA. We use the Synopsys VCS 2021.09 simulation tool in the Verification process which involves deploying test cases to verify these assertions.
Whenever an assertion is rectified based on the prompt given to the LLM, we subsequently re-verify SVA, checking if any ambiguity or errors exist using the logs of the SVA simulation tool. If any error is identified by the simulator, the LLM is fed with error; based on the message in the logs of the simulation tool, the LLM is prompted to analyze, understand, rectify, and generate new assertions. This process involves a human in the loop to feedback and refine prompts.

\begin{figure}[htb]
	\centering
	\captionsetup{justification=centering}
	\includegraphics[width=1.0\linewidth]{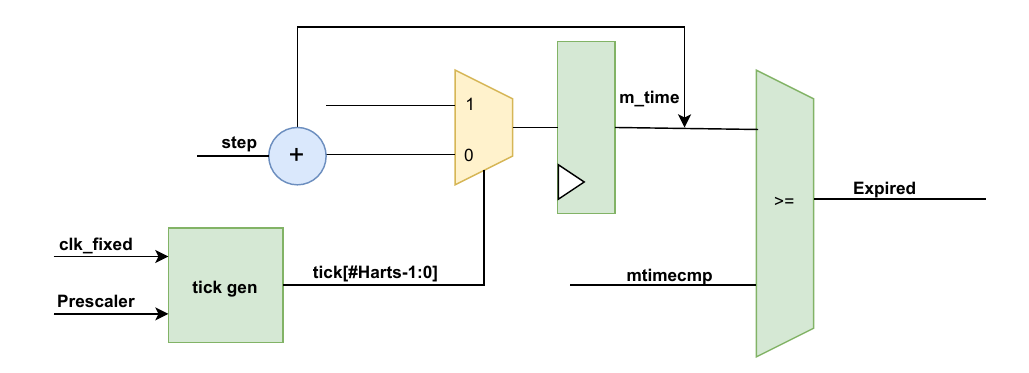}
	\caption{Block Diagram of \texttt{RV Timer}}
	\label{fig:block-rv-timer}
\end{figure}

\subsection{Case study} We conducted our preliminary experiments on an OpenTitan design, \texttt{RV Timer}. The \texttt{RV Timer} has \texttt{clk\_fixed} as the fixed clock input and \texttt{prescaler} that divides the input clock frequency to a lower frequency. This lower frequency is later used by the timer. The timer's count is incremented based on timer ticks generated, which is determined by the \texttt{prescaler} value. The generated ticks are sent to multiple hardware threads. A step value serves as an input to the timer, determining the increment step for each tick. The \texttt{mtime} register is the main timer counter that gets incremented with each tick and holds the current tick value. The comparison to check if a timer has reached a specific time value is performed using the \texttt{mtimecmp} register. The output signal expired is triggered when the \texttt{mtime} value is greater than or equal to the \texttt{mtimecmp} register value. Fig . \ref{fig:block-rv-timer} shows the block diagram of the \texttt{RV Timer}.
Some raw assertions of \texttt{RV Timer} generated by LAAG-RV are:
\begin{lstlisting}
	
	(*@\textbf{Assertion 1:}@*) 
	// Check if tick_count resets to zero when the module is reset or when not active.
	property tick_count_reset; 
	(@(posedge clk_i) disable iff (!rst_ni) (!active) |-> (tick_count == 0));
	end property
	assert property(tick_count_reset);
	
	(*@\textbf{Assertion 2:}@*)
	//Verify that a tick is generated correctly when tick_count reaches prescaler.
	property tick_generate;
	(@(posedge clk_i) disable iff (!rst_ni)(tick_count >= prescaler && active) |=> tick);
	end property
	assert property(tick_generate);
	
	(*@\textbf{Assertion 3:}@*)
	//Ensure tick_count increments correctly unless it matches the prescaler.
	
	property tick_count_increment;
	(@(posedge clk_i) disable iff (!rst_ni || !active) (tick_count < prescaler) |=> (tick_count == $past(tick_count) + 1));
	end property
	assert property(tick_count_increment);
	
\end{lstlisting}

Some of the above LLM-generated assertions have errors. For example, the above { \bf Assertion 1} is incorrect, as identified through simulation. When the log file is supplied as a prompt to the LLM,  \textcolor{black}{it reported the possible causes of the issue, which might be incorrect signal behavior, incorrect assertion timing, or an initial condition not being met. Subsequently, when the LLM is prompted to fix the issues, the LLM finds the timing issue and adds a delay condition to the assertion.}
\textcolor{black}{In the first training prompt itself, the LLM identified the error's cause as a timing issue in the assertion 1 and corrected it, later verified by the simulation tool. The LLM outputted the correct assertion as:}

\begin{lstlisting}
	(*@\textbf{Modified Assertion 1:}@*) 
	// Check if tick_count resets to zero when the module is reset or when not active.
	property tick_count_reset;
	( @(posedge clk_i) disable iff (!rst_ni) (!active) |-> ##1 (tick_count == 0));
	end property
	assert property(tick_count_reset);
\end{lstlisting}

This modification introduces a one-clock cycle delay (\#\#1) before checking if tick\_count equals zero, accommodating any synchronization or propagation delays. Some assertions require multiple prompts due to multiple errors, such as missing signals. The {\bf Assertion 2}, for example, is incorrect in its raw form. The LLM reviewed the RTL for tick generation logic as its primary check, and also reviewed the synchronization conditions and modified the assertion. However, the modified assertion still failed in the simulation run. After three iterations, the LLM could generate a correct assertion upon re-prompting LLM. The modified assertion is shown below.

\begin{lstlisting}
	(*@\textbf{Modified Assertion 2:}@*) 
	
	//Verify that a tick is generated correctly when tick_count reaches the prescaler.
	
	property tick_generate;
	(@(posedge clk_i) disable iff (!rst_ni) (tick_count >= prescaler && active) |-> tick);
	end property
	assert property(tick_generate);
\end{lstlisting}
The modified assertion uses $|->$ (overlapping implication) to check for a tick only in the next cycle immediately after the condition is true. 
While {\bf Assertion 3} shown above is correct in its raw form itself. Therefore, no further prompting of LLM needs to be done. Sometimes, a wrong output may be encountered from the test cases. In such instances, it can be a potential cause for the issue with the design implementation, and manual inspection of the design implementation is required. LLM has generated a few more assertions, as shown below. Some of these assertions are modified from their raw form before successfully passing the simulation runs. However, it is not guaranteed that the assertions generated are enough to cover all the design aspects. More assertions may be required to cover the complete design.

\begin{lstlisting}
	(*@\textbf{Assertion 4:}@*) 
	
	//Assertion to check that tick_count resets when it matches prescaler.
	property tick_count_prescaler;
	(@(posedge clk_i) disable iff (!rst_ni || !active)(tick_count == prescaler) |=> (tick_count == 0));
	end property
	assert property(tick_count_prescaler);
	
	(*@\textbf{Assertion 5:}@*)
	
	//Assertion to check that mtime_d updates correctly based on the value of step.
	property update_mtime_d;
	(@(posedge clk_i) (mtime_d == mtime + step));
	end property
	assert property(update_mtime_d);
	
	(*@\textbf{Assertion 6:}@*)
	
	//Assertion to verify that each interrupt signal is asserted correctly when mtime exceeds or meets mtimecmp.
	generate
	for (genvar t = 0; t < N; t++) begin : gen_intr_assert
	property (mtime_exceed);
	@(posedge clk_i) disable iff (!rst_ni || !active)
	(mtime >= mtimecmp[t]) |=> intr[t] );
	end property
	assert property(mtime_exceed);
	end
	endgenerate
	
	
	(*@\textbf{Assertion 7:}@*)
	
	//Assertion to checks that interrupts are only asserted when the timer is active.
	generate
	genvar t;
	for (t = 0; t < N; t++) begin : gen_intr_assert_t 
	property (interrupt_assert);
	@(posedge clk_i) disable iff(!rst_ni)(active && (mtime >= mtimecmp[t])) |=> intr[t]);
	end property
	assert property(interrupt_assert);
	end
	endgenerate
	
	
\end{lstlisting}

\section{Experimental Results and Discussion}

We have used OpenAI's GPT-4 \cite{openai2023gpt4} LLM to generate the SVA. The capacity of the model stands at 1280000 tokens of the context window.  Our framework is evaluated on various designs from the OpenTitan \cite{Opentitan} repository. We have used Synopsys VCS tool check the generated SVA during ABV.

\subsection{Results of LAAG-RV}
\textcolor{black}{Table \ref{performance_report} shows the performance report of our proposed framework LAAG-RV. The first column, ``Module" shows various designs used for our experiments. The second column, ``OT Assert." shows the number of SVA provided in OpenTitan, and the third column, ``LAAG-RV Assert." shows the number of SVA generated using our framework, ``VCS Sim Time" denotes the simulation time.} 

The preliminary raw assertions generated contain flaws, particularly syntax errors, which are flagged by the simulation tool. To address these errors, the relevant error messages from the log file were manually fed into LLM to correct the assertions. Interestingly, the LLM identified and corrected most of these errors within a single prompt, enabling the assertions to pass the test cases. Additionally, during the prompting process, LLM identified errors and recognized missing or overlooked signals, subsequently incorporating them into the generated assertions.
Similarly, LLM could identify timing-related error and generate the extra SVA later, which passes all test cases. We observed that LLM provided information about the assertion's issues, the potential reason for the error,  and guidelines to fix the errors. This helps to identify the main reason for the failure and craft appropriate prompts.

\begin{table}[t]
	\begin{center}
		\scriptsize
		\caption{Performance Report}
		\begin{tabular}{|c|c|c|c|c|c|c|}
			\hline
			\textbf{Module} & \textbf{OT} & \textbf{LAAG-RV} & \textbf{ChIRAAG} & \textbf{\#Common} & \textbf{VCS}  \\
			& \textbf{Assert.} & \textbf{Assert.} & \textbf{Assert.} &  \textbf{Assert.} & \textbf{Sim}   \\
			&&&&& \textbf{Time}\\
			&&&&& \textbf{(ns)}\\
			\hline
			\texttt{RV Timer} & 0 & 7 & 11 & 5 & 80 \\
			\hline
			\texttt{PattGen} & 0 & 7 & 9 & 4 & 110 \\
			\hline
			\texttt{GPIO} & 0 & 6 & 8 & 4 & 190 \\
			\hline
			\texttt{ROM\_Ctrl} & 6 & 14 & 11 & 7 & 250 \\
			\hline
			\texttt{sram\_ctrl} & 0 & 11 & 14 & 9 & 100 \\
			\hline
			\texttt{adc\_ctrl} & 5 & 7 & 8 & 5 & 460 \\
			\hline
		\end{tabular}
		\label{performance_report}
	\end{center}
\end{table}

Interestingly, the LLM produces more assertions than those originally provided for the OpenTitan design. For instance, while the HMAC design includes only 4 assertions, the LLM generates 12. These LLM-generated assertions are crucial as they cover aspects of the design overlooked by the OpenTitan assertions. For example, one of the assertions is:

\begin{lstlisting}
	property HMAC_process_start;
	@(posedge clk) disable iff (rstt_n)
	(msg_fifo_reqq && hmac_ena) |-> ##1 reg_hash_startt)
	else $fatal("Assertion failed");
	end property
	assert property HMAC_process_start;
\end{lstlisting}

It is essential to verify the HMAC process start property. The process should begin whenever there is a request in the message FIFO and the HMAC signal is enabled. Specifically, if \texttt{msg\_fifo\_reqq} and \texttt{hmac\_ena} are true, then the signal \texttt{req\_hash\_startt} must be true on the next rising edge of the clock, unless a reset (\texttt{rstt\_n}) is active. If this condition is violated, an error message is triggered. In addition, for the \texttt{ROM Controller}, the LLM generated the following SVA, which is not included in the provided OpenTitan assertions:

\begin{lstlisting}
	(*@\textbf{Assertion 8:}@*) 
	property Checker_done;
	@(posedge clk) (checkerr_done && (p !== 'Checking' && p !== 'Done')) |-> ##1 (p === 'Invalid');
	assert property Checker_done;
	
	(*@\textbf{Assertion 9:}@*) 
	property Counter_done;
	@(posedge clk) (counterr_done |-> (p !== 'Reading Low' && p_next !== 'Invalid')
	assert property Counter_done;
	
	(*@\textbf{Assertion 10:}@*)
	property Current_state;
	@(posedge clk) (((p !== 'Invalid') && rstt_ni && kmac_rom_last) |-> ##1 kmac_rom_vld)
	assert property Current_satate
\end{lstlisting}

The above \textbf{Assertion 8} verifies that when the system state (\texttt{p}) is neither in the \texttt{Checking} nor \texttt{Done} states, the signal \texttt{checkerr\_done} should not be true. If this condition is violated, the state \texttt{p} must transition to \texttt{Invalid} in the next clock cycle. \textbf{Assertion 9}  ensures that if the \texttt{counterr\_done} signal is true, the state \texttt{p} should either not be \texttt{Reading Low} in the current state or should not transition to \texttt{Invalid} in the next state. \textbf{Assertion 10} states that if the current state \texttt{p} is not \texttt{Invalid}, and both the reset condition \texttt{rstt\_ni} and the signal \texttt{kmac\_rom\_last} are true, then the signal \texttt{kmac\_rom\_vld} should be true in the next clock cycle.

This demonstrates the usefulness of LLMs in generating assertions for the ABV of designs. However, it should be noted that LLM-generated assertions still require manual verification to ensure their correctness. In some cases, repeated prompting is necessary to achieve self-correction and refine the assertions.

Overall, LLMs are showcasing their promise to revolutionize verification. 

%by automating  generation of correct and consistent assertions. 
\begin{figure}[htbp]
	\centering
	\includegraphics[width=0.4\textwidth]{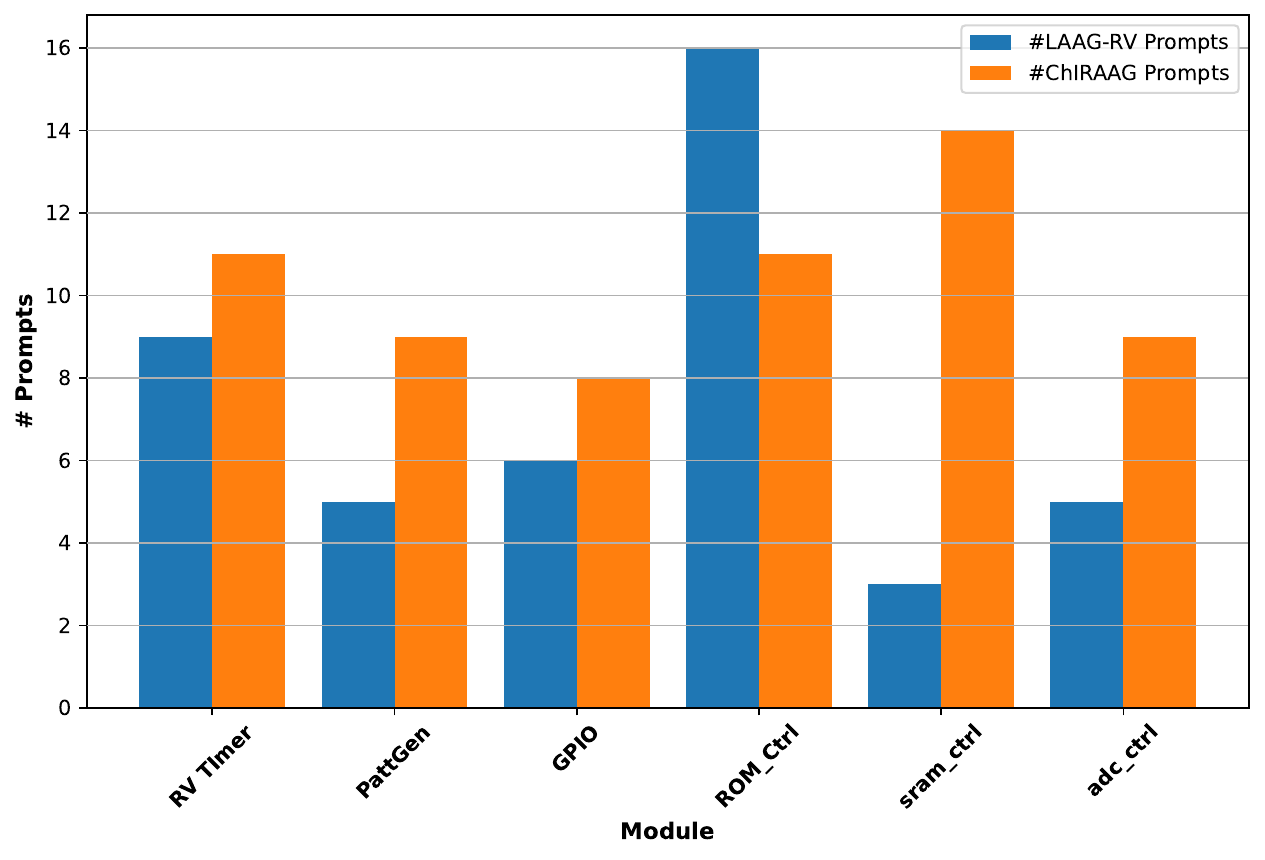}
	\caption{Comparison between the number of LAAG-RV and ChIRAAG Prompts}
	\label{fig:compare}
\end{figure}

\subsection{Comparison with ChIRAAG}
\textcolor{black}{ 
	Our proposed framework is compared with the ChIRAAG \cite{mali2024chiraag}, which follows a different strategy of generating SVA but with a similar goal. In Table \ref{performance_report},   ``ChIRAAG Assert." shows the number of SVA generated using ChIRAAG. ``\#Common Assert" denotes the number of functionally common assertions generated in ChIRAAG and LAAG-RV. It may be noted that ChIRAAG generates more assertions than LAAG-RV since it takes advantage of structural formatting of specification in JSON and also a step-by-step error handling mechanism in each case. However, creating JSON formatting is manual which is avoided in LAAG-RV by a one-time loop of Verilog implementation fed to the LLM before generating the SVA. Interestingly, most of the assertions generated by them are common which shows the consistency of LLMs in generating assertions from the specification.} For example:

\begin{lstlisting}
	(*@\textbf{Assertion 11:}@*) 
	//Assertion generated using LAAG-RV for checking tick_count behaviour
	property tick_count_increment;
	@(posedge clk_i) disable iff (!rst_ni || !active) (
	tick_count < prescaler) |=> (tick_count == $past(tick_count) + 1);
	end property
	assert property (tick_count_increment);
	
	(*@\textbf{Assertion 12:}@*) 
	
	//Assertion generated using ChIRAAG for checking tick_count behaviour
	property tick_count_increment;
	@(posedge clk_i)
	(rst_ni && active && (tick_count < prescaler)) |=> (
	tick_count == $past(tick_count) + 1);
	endproperty
	assert property (tick_count_increment);
	
\end{lstlisting}

\textcolor{black}{Functionally, both the above assertions achieve the same goal of ensuring \texttt{tick\_count} increments by 1 under the specified conditions. The difference lies in their approach to evaluation. {\bf Assertion 11}, which is generated using LAAG-RV Uses \texttt{disable iff} to prevent the property from being evaluated when \texttt{rst\_ni} or \texttt{active} are low. {\bf Assertion 12}, which is generated using ChIRAAG, always evaluates the property but only asserts the behavior when \texttt{rst\_ni} is high, \texttt{active} is high, and \texttt{tick\_count < prescaler}. Despite these differences, both assertions ensures the increment behavior of \texttt{tick\_count}. In many other cases, we have encountered such variations where functionality is similar but representations are different.}

Fig. \ref{fig:compare} shows the comparison of the number of prompts required by LAAG-RV and by ChIRAAG, respectively. LAAG-RV takes relatively less number of iterations as compared to ChIRAAG for most designs because of the one-time loop of Verilog code fed to the LLM before generating the SVA. This step ensures the signal names are synchronized with the generated SVA signal names, thereby overcoming missing signal errors in fewer iterations.

\section{Conclusion}
In this work, we have developed a framework LAAG-RV to generate SVA from the natural language specification. This framework operates with the help of a one-time Verilog loop, which is helpful in generating correct signal names and is the main reason for rectifying SVA errors in fewer iterations. We noticed that some of the initially generated assertions using our framework have issues; based on manual prompting using the error of a simulator, the LLM was able to refine the SVA. Providing initial knowledge to the LLM by feeding it with domain-related documents makes it produce more efficient and correct assertions in fewer steps.  
This study focused on the correct assertion generation from English language specifications of a design. In future, we will explore LLMs to check the other two Cs, i.e., the consistency and completeness of assertions with respect to the specification.

\renewcommand*{\bibfont}{\small}
\bibliographystyle{IEEEtran}
\vspace{-0.1 cm}
\bibliography{ref.bib}

\end{document}